# Three-level description of the domino cellular automaton

*Zbigniew Czechowski and Mariusz Białecki*

Institute of Geophysics Polish Academy of Sciences, 01-452 Warsaw,

Księcia Janusza 64, Poland, zczech@igf.edu.pl, bialecki@igf.edu.pl

**Abstract**

Inspired by the approach of kinetic theory of gases, a three-level description (microscopic, mesoscopic and macroscopic) of cellular automaton is presented. To provide an analytical treatment a simple domino cellular automaton with avalanches was constructed. Formulas concerning exact relations for density, clusters, avalanches and other parameters in an equilibrium state were derived. It appears that some relations are approximately valid for deviations from the equilibrium, so the adequate Ito equation could be constructed. The equation provides the time evolution description of some variable on the macroscopic level. The results also suggest a motive for applying of the procedure of construction of the Ito equation (from time series data) to natural time series.

## 1. Introduction

Cellular automata replace equations of physics by using some rules concerning cells behavior on some grid which models a medium. They are very handy for computer simulations. They allow to analyze discrete systems in which nonlinear complex interactions can be easily described by simple rules. In this way we omit numerical procedure of discretization of differential equations. Moreover, they are useful tools of modeling in which it is easy to implement a stochastic contribution.

Cellular automata can be constructed as toy models of various complex phenomena, whose physical nature is difficult to understand and explain on the ground of physical equations e.g., phase transitions, SOC, turbulence. Being the discrete models, cellular automata naturally generate discrete time series. Methods of time series analysis and modeling can be tested by using these data. Here, a very important problem appears: for complex phenomenon the recorded time series may be treated as a realization of some stochastic process. Therefore, we are looking for an adequate stochastic model, which could reconstruct this process. However, the founded model describes often macroscopic (observable) behavior of the phenomenon under investigation. How to link this macroscopic description with the microscopic structure and rules of the input cellular automaton? The main aim of the present paper is to find and explain these relations.

We present the three-level description of a phenomenon: macroscopic, mesoscopic and microscopic. Excellent example of such approach is the description of gases (see e.g., [1]). The microscopic description is composed of a huge system of Newton equations for gas molecules. The basic object on the mesoscopic level is the velocity distribution function of molecules governed by the integral-differential nonlinear Boltzmann equation. The macroscopic description is given by moments of the distribution function: gas density, macroscopic velocity, pressure and temperature of the gas. By using the procedure of taking successive moments of the Boltzmann equation, the system of macroscopic equations is derived for a continuous fluid medium: the continuity equation, the Navier-Stokes equation and the heat transport equation. In such a description the higher levels have a clear interpretation on the lower levels.

In our paper we make similar operations. Instead of a gas we investigate a simple 1-D stochastic cellular automaton with avalanches, whose physical motives may correspond, for example, to processes on the tectonic fault. Therefore, slow filling of grid cells may be interpreted as a slow pumping of the elastic energy to the fault region, and avalanches – as a rapid energy release during earthquakes. The microscopic description of the automaton is given by its geometry and rules. Mesoscopic level is composed of equations for the cluster size distribution function and the macroscopic level – of equations for moments of this distribution, including the Ito equation describing the evolution of density of occupied cells on the grid.

In this way we perform sort of a reversed approach than those usually employed in the field of cellular automata - we derive nonlinear equations governing the behavior of the cellular automaton. Advantages are essential: we find clear dependences between macroscopic variables and microscopic states on the grid. Usually, in the majority of papers the macroscopic behavior of automata has been detected by a large number of computer simulations. In fever papers one can find some dependences between different levels of description and the derived macroscopic laws. They are exact in the case of deterministic automata (e.g., [2], [3]), or approximate as the mean field theory, for automata with a stochastic contribution (e.g., [4] – [8]). In their new paper [9] the authors extend the mean field approximation of BTW model to one that includes the correlation between pairs of nearest neighbors.

In our paper we derive exact relations and solutions for the simple 1-D domino cellular automaton at the equilibrium state. However, for an evolution of deviations from the equilibrium the derived Ito equation is yet some approximation. This equation may compose also a good, individual model of a phenomenon on the macroscopic level if we have, for a given phenomenon, only a time series of some observable. Here we have assumed that this observable is the density of occupied cells on the grid.

The Ito equation describes an evolution of diffusion Markov process. However, for real, e.g., geophysical time series, the question arises; whether the complex phenomenon under investigation may be reliably described by the diffusive Markov process. The present paper shows that at any rate for time series generated by the domino cellular automaton (which is a toy model of a complex phenomenon) the Ito equation is a satisfactory model. This result suggests a motive for applying of the procedure of construction of the Ito equation ([10] - [12]) to natural time series.

The plan of the paper is as follows. In section 2, we introduce the domino cellular automaton and derive exact equations and relations for the equilibrium state. In section 3, we find formulas for cluster size distribution for deviation from the equilibrium. Using this formulas we derive, in section 4, the Ito equation for the density of occupied cells on the grid. Two appendixes (A and B) contain some mathematical details of the derivation. In section 5 some concluding remarks are presented.

## 2. Domino cellular automaton

The generalized domino automaton is introduced in our parallel paper [13], however here it is useful to present the complete derivation of equilibrium equations for a simplified case.

**The rules**

In this simple cellular automaton, particles are added and lost from a 1-D grid according to the procedure:

at each time step a particle is randomly added to one of the boxes,

- when it hits an empty box the box becomes the occupied one,
- when it hits an occupied box the whole cluster (i.e., the chain of neighboring occupied boxes) is lost; the avalanche appears (see Fig. 1).

We will characterize the (average) state of the model by using the variable $\rho$ which denotes the rate of number of occupied boxes to the grid size (or the probability that the given box is occupied). The variable ρ increases uniformly between avalanches and drops violently in an avalanche. After an initial saturation stage, ρ fluctuates around the equilibrium state $\rho_{eq}$ (see Fig. 2).

**Three level description**

We can distinguish three levels of description of the grid state:

- microscopic – given by rules of the automaton and states of each grid box (occupied or empty),
- mesoscopic – given by cluster size distribution functions $n_i$ (occupied clusters) and $n^0_i$ (clusters of empty boxes), where $i$ is the cluster size,
- macroscopic – given by moments (or other averages) of these distribution functions.

It is easy to find the following relations: the total number (per box) $n$ of clusters is given by the formula (the zeroth moment of $n_i$):

$$n = \sum_{i\geq 1} n_i = \sum_{i\geq 1} n_i^0 , \tag{1}$$

ρ is the first moment of $n_i$

$$\rho = \sum_{i\geq 1} n_i \, i , \tag{2}$$

and therefore:

$$1-\rho = \sum_{i\geq 1} n_i^0 i . \tag{3}$$

The derivation of the balance equations will be provided on the macroscopic and mesoscopic levels.

In the text below we use the designation “cluster” for the occupied cluster.

**The balance for ρ**

The probability that the particle hits an empty box should be balanced by the probability that it hits an occupied box and triggers an avalanche of any size:

$$1-\rho=\sum_{i\geq 1} n_i i^2 \quad \text{or} \quad 1-\rho=\rho\sum_{i\geq 1} w_i i \ , \tag{4}$$

where the probability of avalanche of size $i$ is given by $w_i = in_i/\rho$, because hitting any of boxes of $i$-cluster triggers the event.

**The balance for *n***

***Gains:*** a new cluster will appear if the particle hits any of $i$-2 boxes from the 'interior' (i.e., from empty $i$-cluster without two boxes on its ends) of empty cluster of size $i \geq 3$.
***Losses:*** one cluster disappears if particle hits any cluster (hits any occupied box) or if it hits a single empty box between two clusters (i.e., hits an empty 1-cluster; therefore, these two clusters link).

Therefore:

$$\sum_{i\geq 3} n_i^0 (i-2)=\rho+n_1^0 , \tag{5}$$

or using (1) and (3)

$$2\rho+2n=1. \tag{6}$$

The two balance equations, (4) and (6), show simple relations between three moments of $n_i$:
$m_0=\sum_{i\geq 1} n_i = n$, $m_1=\sum_{i\geq 1} n_i i=\rho$ and $m_2=\sum_{i\geq 1} n_i i^2$ ($m_2$ relates to the average avalanche size; $m_2=\langle w\rangle\rho$) in the equilibrium state:

$$m_1+m_2=1 \qquad\qquad m_2=1-\rho \tag{7}$$

or

$$2m_0+2m_1=1 \qquad\qquad m_0=\frac{1}{2}-\rho . \tag{8}$$

Unfortunately, this is not a closed system of equations, therefore we should derive balance equations for $n_i$.

**The balance for $n_i$, $i \geq 3$**

***Losses:*** a new dropping particle may decrease the number of $i$-clusters if it hits one of two empty boxes on the perimeter of the cluster or it hits one of $i$ boxes of the cluster (triggering the avalanche):

$$losses=(i+2)n_i . \tag{9}$$

***Gains:*** a dropping particle may increase the number of $i$-clusters if:

a) it hits one of two boxes on the perimeter of a ($i$-1)-cluster (but not causing the linking with the neighboring cluster; i.e., if the adjacent empty cluster is of size bigger than 1):

$$gains(a) = 2n_{i-1}\sum_{i\geq 2}\frac{n_i^0}{n^0} = 2n_{i-1}\left(1-\frac{n_1^0}{n}\right). \tag{10}$$

b) it hits an empty 1-cluster separating $k$-cluster and ($i$-1-$k$)-cluster:

$$gains(b) = n_1^0\sum_{k=1}^{i-2}\frac{n_k n_{i-1-k}}{n^2}. \tag{11}$$

Therefore, the balance equation for $i$-clusters (for $i \geq 3$) is as follows:

$$(i+2)n_i = 2n_{i-1}\left(1-\frac{n_1^0}{n}\right) + n_1^0\sum_{k=1}^{i-2}\frac{n_k n_{i-1-k}}{n^2}. \tag{12}$$

**The balance for $n_1$ and $n_2$**

There are no term *gains*($b$) for $n_2$, so

$$4n_2 = 2n_1\left(1-\frac{n_1^0}{n}\right). \tag{13}$$

Gains for the number of 1-clusters are proportional to the number ($i$-2) of 'interior' boxes in empty $i$-clusters of the length ≥ 3:

$$gains = \sum_{i\geq 3}(i-2)n_i^0 = \sum_{i\geq 1}(i-2)n_i^0 + n_1^0 = (1-\rho) - 2n + n_1^0, \tag{14}$$

and therefore the balance equation has the form

$$3n_1 = (1-\rho) - 2n + n_1^0. \tag{15}$$

It is easy to check that after summing up left and right sides of balance equations (13), (15) and (12) for $i \geq 3$, formula (6) is reconstructed (terms with $n^0_1$ vanish). However, to complete equations (12), (13) and (15) to the closed form we derive the balance equation for empty 1-clusters $n^0_1$.

**The balance for $n^0_1$**

***Gains:*** a new empty 1-cluster will be created if a particle hits one of two boxes (located one empty box away of the neighboring occupied cluster) of any empty $i$-cluster, $i \geq 2$:

$$gains = 2\sum_{i\geq 2}n_i^0 = 2n - 2n_1^0. \tag{16}$$

***Losses:*** the empty 1-cluster will vanish if a particle hits the empty 1-cluster or hits any cell of the occupied cluster adjoining to the empty 1-cluster:

$$losses = n_1^0 + 2n_1^0 \sum_{i\geq 1} \frac{n_i i}{n} = n_1^0 \left(1 + 2\frac{\rho}{n}\right). \tag{17}$$

Therefore, the balance equation has the following form:

$$n_1^0 (3n + 2\rho) = 2n^2. \tag{18}$$

Balance equations for empty clusters of any size were also derived and they are introduced in our parallel paper [13].

The obtained closed system of equations (12), (13), (15) and (18) may be treated as recursive formulas for $n_i(\rho)$:

$$n_1 = \frac{1}{3}\left(1 - \rho - 2n + n_1^0\right), \tag{19}$$

$$n_2 = \frac{1}{2}\left(1 - \frac{n_1^0}{n}\right) n_1, \tag{20}$$

$$n_i = \frac{1}{i+2}\left[2\left(1 - \frac{n_1^0}{n}\right) n_{i-1} + n_1^0 \sum_{k=1}^{i-2} \frac{n_k n_{i-1-k}}{n^2}\right] \quad \text{for } i \geq 3, \tag{21}$$

where, by using (6) and (18), $n$ and $n^0{}_1$ can be given in the equilibrium state in terms of $\rho$ by:

$$n = \frac{1}{2} - \rho, \tag{22}$$

$$n_1^0 = \frac{(1-2\rho)^2}{3 - 2\rho}. \tag{23}$$

The equilibrium value of $\rho_{eq}$ = 0.3075 was calculated numerically from equation (2), where $n_i$ were given by (19) - (21). Then, we can calculate other equilibrium values of moments of $n_i$:

the average number of clusters (per cell): $< n >= m_0 = 0.1925$,

the average cluster size (per cell): $< i >= \frac{m_1}{m_0} = 1.597$,

the average avalanche size (per cell): $<w>=\frac{m_2}{m_1}=2.252.$

The formulas (19) - (23) for $n_i(\rho)$, $n(\rho)$ and $n^0{}_1(\rho)$ were derived from balance equations and, therefore, they are valid only for the equilibrium value of $\rho = \rho_{eq}$ (see Fig. 6 for $<n>(\rho)$, $<i>(\rho)$, $<w>(\rho)$).

## 3. Formulas for $n_i(\rho)$ for deviation from the equilibrium

The mechanism of the domino cellular automaton leads to different cluster distribution than in the 1-D percolation. However, some relations are maintained. For the 1-D percolation the cluster size distribution has the geometric form:

$$n_k(\rho)=\rho^k(1-\rho)^2. \tag{24}$$

Figure 3 shows that for the equilibrium value $\rho = \rho_{eq}$, a similar geometric distribution is valid also for our automaton for $k \geq 3$. Moreover, ratios $n_{k+1}/n_k$ calculated from (19) – (21) are nearly linear functions of $\rho$ in the range $0.23 < \rho < 0.4$ under investigation. We observe that the ratios become nearly identical for $k \geq 3$. These observations suggest the following hierarchy for $n_i(\rho)$:

$$\begin{aligned}
&n_1(\rho)=(1-\rho)^2 a_1(\rho)=(1-\rho)^2 a_1\rho,\\
&n_2(\rho)=n_1(\rho)a_2(\rho),\\
&n_3(\rho)=n_2(\rho)a_3(\rho),\\
&n_k(\rho)=n_{k-1}(\rho)a_4(\rho)^{k-3} \quad \text{for } k\geq 4,
\end{aligned} \tag{25}$$

where $a_i(\rho)$, $i$ = 1, 2, 3, 4 are linear functions of $\rho$. Functions $a_2(\rho)$, and $a_4(\rho)$ can be found from the Taylor expansion of the ratios: $n_2(\rho)/n_1(\rho)$ and $n_4(\rho)/n_3(\rho)$ in $\Delta\rho = \rho - \rho_{eq}$ around the equilibrium value $\rho_{eq}$:

$$\frac{n_2(\rho)}{n_1(\rho)}=\frac{-1-2\rho_{eq}}{2(-3+2\rho_{eq})}+\frac{4\Delta\rho}{(-3+2\rho_{eq})^2}+O[\Delta\rho^2], \tag{26}$$

$$\frac{n_4(\rho)}{n_3(\rho)}=\frac{-17-32\rho_{eq}-36\rho_{eq}^2-80\rho_{eq}^3}{3(-3+2\rho_{eq})(7+8\rho_{eq}+20\rho_{eq}^2)}+\frac{2(251+8\rho_{eq}+3016\rho_{eq}^2+2080\rho_{eq}^3+2480\rho_{eq}^4)\Delta\rho}{3(-3+2\rho_{eq})^2(7+8\rho_{eq}+20\rho_{eq}^2)^2}+O[\Delta\rho^2], \tag{27}$$

which gives for $\rho_{eq} = 0.3075$:

$$a_2(\rho)=0.703\rho+0.122, \tag{28}$$

$$a_4(\rho)=0.560\rho+0.233. \tag{29}$$

Next, we calculate the constant $a_1$ by using the equation:

$$a_2(\rho)a_3(\rho)=\frac{\dfrac{\rho}{n_1(\rho)}-1-2a_2(\rho)}{n_2(\rho)\dfrac{3-2a_4(\rho)}{(1-a_4(\rho))^2}}=\frac{\dfrac{\frac{1}{2}-\rho}{n_1(\rho)}-1-a_2(\rho)}{\dfrac{1}{1-a_4(\rho)}}=a_2(\rho)a_3(\rho), \tag{30}$$

where the left (right) hand side results from the formula for $\rho=\sum_{i\geq1} in_i$ ($n=\sum_{i\geq1} n_i=\frac{1}{2}-\rho$) and denominators are equal to the sums $\sum_{i\geq3} ia_4^{\,i-3}$ and $\sum_{i\geq3} a_4^{\,i-3}$ respectively. For $\rho=\rho_{eq}$ we obtain $a_1= 0.837$ or

$$a_1(\rho)=0.837\rho. \tag{31}$$

At the end we find $a_3(\rho)$ by the Taylor expansion of

$$\frac{\rho-n_1(\rho)-2n_2(\rho)}{n_2(\rho)\dfrac{3-2a_4(\rho)}{(1-a_4(\rho))^2}}=a_3(\rho)+O[\Delta\rho^2], \tag{32}$$

which for $\rho_{eq}=0.3075$ leads to

$$a_3(\rho)=1.4225\rho+0.048. \tag{33}$$

The obtained ratios $a_2(\rho)$, $a_3(\rho)$, $a_4(\rho)$ fit well (see Fig. 4) to the simulation results.

The presented method enables us to derive from (19) – (21) satisfactory formulas for $n_i(\rho)$ as functions of $\rho$ although the expressions (19) – (21) are valid for $\rho=\rho_{eq}$ only. However, it appears that deviations of these expressions from the real behaviour have a systematic character and, therefore, these deviations are reduced by using the presented procedure.

It should be underlined that the expressions (25) with (28), (29) and (33) for $\rho=\rho_{eq}$ are equal to adequate formulas (19) – (21) due to employing the Taylor expansions around $\rho_{eq}$.

Having satisfactory formulas for $n_i(\rho)$ (see Fig. 5) we can derive expressions for averages $n(\rho)$, $\langle i\rangle(\rho)$ and $\langle w\rangle(\rho)$ as functions of $\rho$ :

$$n(\rho)=\sum_{i\geq1} n_i(\rho)=(1-\rho)^2 a_1(\rho)\left\{1+a_2(\rho)\left[1+\frac{a_3(\rho)}{1-a_4(\rho)}\right]\right\}, \tag{34}$$

$$\langle i\rangle(\rho)=\frac{\sum_{i\geq1} i\, n_i(\rho)}{\sum_{i\geq1} n_i(\rho)}=\frac{\rho}{n(\rho)}, \tag{35}$$

$$< w > (\rho) = \frac{\sum_{i\geq 1} i^2 n_i(\rho)}{\sum_{i\geq 1} i n_i(\rho)} = (1-\rho)^2 a_1 \left\{ 1 + a_2(\rho)\left[ 4 + a_3(\rho)\frac{9 - 11a_4(\rho) + 4a_4^{\,2}(\rho)}{[1-a_4(\rho)]^3} \right] \right\}. \quad (36)$$

Figure 6 shows that these expressions fit the simulation results very well.

## 4. The Ito equation for the domino cellular automaton

We are going to derive the Ito equation for the variable ρ, which may be treated as a Markov diffusion process ρ(*t*). We have assumed the **convention in the automaton that what provokes us to the investigation of the grid state is the avalanche.** Therefore, **we monitor avalanche sizes and states of the automaton after each avalanche**. As a result we obtain time series of avalanche sizes *w*(*i*) and densities ρ(*i*), *i* = 1, 2, … . By using this convention we avoid less interesting stairs-like increase of ρ(*i*) in periods between avalanches.

In order to derive functions *a*(ρ) and *b*(ρ) in the Ito equation we need the transition probability *P*(ρ'|ρ) (see [14]).

According to our convention the effective change of ρ is a result of growth of ρ in stairs before an avalanche starts and a drop of ρ in the avalanche.

Let us derive the probability *EG*(*k*) of effective gain (an increase by *k* boxes) of the number of occupied boxes calculated immediately after an avalanche:

$$EG(k) \equiv P\left( \rho_i + \frac{k}{N}, i+1 \,|\, \rho_i, i \right) = \sum_{s=k+1}^{\infty} (1-\rho)^s \cdot \rho \cdot w_{s-k}(\rho) \quad \text{for } k \geq 0,$$

(37)

where factors under the sum denote:

$(1-\rho)^s$ is the probability of hit one after the other of *s* empty boxes,

ρ is the probability of hit of an occupied box in the next step,

$w_{s-k}(\rho)$ is the probability that this occupied box is a part of the cluster of size *s-k*, and *N* is the number of all boxes in the grid.

Similarly, the probability of effective loss *EL*(*k*) is given by the formula:

$$EL(k) \equiv P\left( \rho_i - \frac{k}{N}, i+1 \,|\, \rho_i, i \right) = \sum_{s=k}^{\infty} (1-\rho)^{s-k} \cdot \rho \cdot w_s(\rho) \quad \text{for } k \geq 1,$$

(38)

where the size of avalanche excels stairs gains before the avalanche. It can be shown (see App. A) that *EG*(*k*) and *EL*(*k*) fulfill the normalization condition:

$$\sum_{k\geq 1} \left[ EG(k) + EL(k) \right] + EG(0) = 1. \quad (39)$$

The functions *a*(ρ) and *b*(ρ) in the Ito equation represent (see Risken 1996) the first and the second moment of the transition probability:

$$a(\rho) \propto \sum_{k\geq 1} k\left[EG(k) - EL(k)\right], \tag{40}$$

$$b(\rho) \propto \sum_{k\geq 1} k^2\left[EG(k) + EL(k)\right]. \tag{41}$$

It is interesting that careful transformations on sums (see App. B) let us to give the expressions for $a(\rho)$ and $b(\rho)$ in terms of ρ and the first and the second moment of $w_k(\rho)$ :

$$a(\rho) \propto \frac{1-\rho}{\rho} - \sum_{k\geq 1} kw_k\,, \tag{42}$$

$$b(\rho) \propto \frac{2-3\rho+\rho^2}{\rho^2} + \frac{2\rho-2}{\rho}\sum_{k\geq 1} kw_k + \sum_{k\geq 1} k^2 w_k\,. \tag{43}$$

Due to the fact that in our automaton a hit of any occupied box belonging to a cluster of size $k$ triggers the avalanche we have

$$\rho \cdot w_k(\rho) = k \cdot n_k(\rho)\,. \tag{44}$$

Then, formulas (42) and (43) may be expressed by the first (i.e., ρ), the second and the third moment of $n_k(\rho)$:

$$a(\rho) \propto \frac{1}{\rho}\left(1-\rho-\sum_{k\geq 1} k^2 n_k\right), \tag{45}$$

$$b(\rho) \propto \frac{1}{\rho}\left\{\frac{1}{\rho}\left[2-3\rho+\rho^2+(2\rho-2)\sum_{k\geq 1} k^2 n_k\right] + \sum_{k\geq 1} k^3 n_k\right\}. \tag{46}$$

Using formulas for $n_i$ derived in Section 3 we can calculate the moments:

$$\sum_{k\geq 1} k^2 n_k = a_1(\rho)(1-\rho)^2\left\{1+a_2(\rho)\left[4+a_3(\rho)\frac{9-11a_4(\rho)+4a_4{}^2(\rho)}{(1-a_4(\rho))^3}\right]\right\}, \tag{47}$$

$$\sum_{k\geq 1} k^3 n_k = a_1(\rho)(1-\rho)^2\left\{1+a_2(\rho)\left[8+a_3(\rho)\frac{27-44a_4(\rho)+31a_4{}^2(\rho)-8a_4{}^3(\rho)}{(1-a_4(\rho))^4}\right]\right\}. \tag{48}$$

In this way we obtain the Ito equation

$$d\rho = a(\rho)dt + \sqrt{b(\rho)}dW(t)\,, \tag{49}$$

where $a(\rho)$ and $b(\rho)$ are known to be drift and diffusion coefficients, respectively, and $W(t)$ is the Wiener process. Functions $a(\rho)$ and $b(\rho)$ are determined explicitly by (45) and (46) with (47) and (48), where parameters $a_i(\rho)$, $i = 1, 2, 3, 4$, are given, in our approximation, by (28), (29), (31) and (33). A comparison of the functions with simulation results is presented in Fig. 7. Clouds of points which illustrate functions $a(\rho)$ and $b(\rho)$ are found by using the histogram method of constructing the Ito equation from time series ([10], [11], [12]). The fit is satisfactory, particularly in the case of function $a(\rho)$. The fit for function $b(\rho)$ is not so excellent due to the deviations in our approximation; $b(\rho)$ is more sensitive, because it is calculated as the second moment of the transition probability. An example of the time series generated by the constructed Ito equation is shown in Fig. 8. Both time series (that given by the automaton (Fig. 2) and that generated by the Ito equation Fig. (8)) are similar; however, due to the excess in the function $b(\rho)$ for $\rho < 0.31$ we observe greater fluctuations of $\rho$ downward in the time series generated by the Ito equation.

Ito equations describe evolution of diffusion Markov processes. The time series $\rho(i)$ can be treated as a realization of some continuous stochastic process $\rho(t)$. Rules of the automaton provide the Markov property for $\rho(i)$. However, it is difficult to prove that the process is a diffusion one, but the results let as ascertain that this may be true.

## 5. Conclusions

The paper presents the three-level description (microscopic, mesoscopic, macroscopic) of the simple cellular automaton with avalanches. The approach was inspired by the success of the kinetic theory of gases which led to the linking of microscopic aspects of molecular dynamics with macroscopic description of the gas as a continuous medium. The theory explained what is the meaning, in a discrete molecular view, of so macroscopic quantities as: velocity, temperature, pressure, viscosity and coefficient of thermal conductivity. It showed how microscopic conservation principles for molecules led to macroscopic equations of hydrodynamics.

Our domino cellular automaton is the discrete model, with a discrete time in addition. The role of the velocity distribution function for the gas fulfills here the cluster size distribution function. Its moments are related to macroscopic quantities: average number of clusters, average density of occupied cells, average cluster size, and average avalanche size. One can also introduce the average square deviation from the average cluster size (i.e., the analog of temperature)

$$< T > = \sum_{i \geq 1} n_i \, (i - < i >)^2 = \rho(< w > - < i >) \,. \tag{50}$$

The model has a self-organizing character; for some $\rho = \rho_{eq}$ it attains the quasi-equilibrium state, in which there are fluctuations of $\rho$ with nearly Gaussian distribution (see our parallel paper [13]) around the equilibrium value of $\rho_{eq}$ (we can call it the Self-Organizing Quasi-Equilibrium; SOQE).

For the equilibrium state we found exact relations between moments and iteration formulas for the cluster size distribution $n_k(\rho_{eq})$. By solving the implicit algebraic equation the equilibrium value of $\rho_{eq}$ was calculated.

For fluctuations of $\rho$ from the equilibrium value approximate formulas for $n_i(\rho)$ were derived. They were used to calculation of averages $n(\rho)$, $<i>(\rho)$ and $<w>(\rho)$ as functions of $\rho$; a good agreement with simulation results was obtained.

Next, the Ito equation, which describes the evolution of density $\rho(t)$, was derived. Functions $a(\rho)$ and $b(\rho)$ are determined by $\rho$ and the second and the third moments of $n_k(\rho)$. Explicit formulas for the moments are given by using approximate formulas for $n_k(\rho)$. Comparison of $a(\rho)$ and $b(\rho)$ with simulation results is satisfactory, particularly for $a(\rho)$. The time series generated by the Ito equation is similar to that given by the domino cellular automaton.

Our approach, although inspired by the kinetic theory of gases, is different from the former one. For example, we can not derive a time-dependent equation for cluster size distribution function (analogous to the Boltzmann kinetic equation). Instead of this we derive the time-dependent Ito equation for $\rho(t)$ (which is treated as a stochastic process). Having $\rho(t)$, the distribution $n_k(\rho(t))$ as well averages $n(\rho(t))$, $\langle i\rangle(\rho(t))$ and $\langle w\rangle(\rho(t))$ can be calculated from formulas (33), (34) and (35). The quantity $\rho$ has a double meaning: it is the first moment of $n_k$, but it also may be treated as an independent quantity, which composes rather the constrain for $n_k$.

Ito equations can constitute useful macroscopic models of complex phenomena, in which microscopic interactions are averaged in an adequate way. They can describe stochastic behavior of some quantities, which are observable (registered as the time series). Then, the histogram procedure of reconstruction of the Ito equation from time series data may be a handy tool of nonlinear time series analysis. Nonlinear modeling of time series are very desirable by many scientists. The histogram procedure fills a gap between two methods: linear stochastic models (ARMA, etc) and nonlinear deterministic models (which lead to deterministic chaos, [15]) - the Ito models are both nonlinear and stochastic. In this context, the derivation of the Ito equation for a simple cellular automaton (which can be a toy model of the complex phenomenon) becomes an inspiration for applying the procedure of reconstruction of the Ito equation for the case of natural time series.

**Acknowledgements**

The work was supported by INTAS 05-1000008-7889 and by Ministry of Science and Higher Education grant no. 30/W-INTAS/2007/0.

**APPENDIX A**

**Checking the normalization condition for** $P\left(\rho_i \pm \frac{k}{N}, i+1 \mid \rho_i, i\right)$.

Because:

$$\sum_{s=k+1}^{\infty}(1-\rho)^s \cdot \rho \cdot w_{s-k} = \sum_{s=1}^{\infty}(1-\rho)^{k+s} \cdot \rho \cdot w_s ,$$

therefore:

$$\sum_{k=1}^{\infty} EG(k) = \rho\left[\sum_{k=1}^{\infty}(1-\rho)^k\right]\left[\sum_{s=1}^{\infty}(1-\rho)^s w_s\right] = \rho\frac{1-\rho}{\rho}\sum_{s=1}^{\infty}(1-\rho)^s w_s .$$

On the other hand:

$$\sum_{k=1}^{\infty} EL(k) = \rho\left[\sum_{k=1}^{\infty}(1-\rho)^{-k}\sum_{s=k}^{\infty}(1-\rho)^s w_s\right] = \rho\left[(1-\rho)^{-1}\sum_{s=1}^{\infty}(1-\rho)^s w_s + (1-\rho)^{-2}\sum_{s=2}^{\infty}(1-\rho)^s w_s + \dots\right] =$$

$$\rho\sum_{s=1}^{\infty}(1-\rho)^s w_s\left\{(1-\rho)^{-1} + (1-\rho)^{-2}\left[1-\frac{(1-\rho)w_1}{\sum}\right] + (1-\rho)^{-3}\left[1-\frac{(1-\rho)w_1}{\sum}-\frac{(1-\rho)^2 w_2}{\sum}\right] + \dots\right\} =$$

$$\rho\sum_{s=1}^{\infty}(1-\rho)^s w_s\left\{\left[(1-\rho)^{-1} + (1-\rho)^{-2} + \dots\right] - \frac{w_1}{\sum}\left[(1-\rho)^{-1} + (1-\rho)^{-2} + \dots\right] - \frac{w_2}{\sum}\left[(1-\rho)^{-1} + (1-\rho)^{-2} + \dots\right] - \dots\right\} =$$

$$\rho\sum_{s=1}^{\infty}(1-\rho)^s w_s\left\{\sum_{k=1}^{\infty}(1-\rho)^{-k}\left(1-\frac{w_1}{\sum}-\frac{w_2}{\sum}-\dots\right)\right\} = -\sum_{s=1}^{\infty}(1-\rho)^s w_s + \sum_{s=1}^{\infty} w_s ,$$

where

$$\sum \equiv \sum_{s=1}^{\infty}(1-\rho)^s w_s$$

Moreover:

$$EG(0) = \rho\sum_{s=1}^{\infty}(1-\rho)^s w_s ,$$

therefore:

$$\sum_{k\ge 1}\left[EG(k) + EL(k)\right] + EG(0) = (1-\rho)\sum_{s=1}^{\infty}(1-\rho)^s w_s - \sum_{s=1}^{\infty}(1-\rho)^s w_s + \sum_{s=1}^{\infty} w_s + \rho\sum_{s=1}^{\infty}(1-\rho)^s w_s = \sum_{s=1}^{\infty} w_s = 1$$

because $w_k$ is the probability.

## APPENDIX B

**Derivation of formulas for *a*(ρ) and *b*(ρ).**

We have:

$$\sum_{k=1}^{\infty} EG(k)k = \rho\left[\sum_{k=1}^{\infty}(1-\rho)^k k\right]\left[\sum_{s=1}^{\infty}(1-\rho)^s w_s\right] = \frac{1-\rho}{\rho}\sum_{s=1}^{\infty}(1-\rho)^s w_s,$$

and:

$$\sum_{k=1}^{\infty} EL(k)k = \rho\left[\sum_{k=1}^{\infty}(1-\rho)^{-k} k\sum_{s=k}^{\infty}(1-\rho)^s w_s\right] = \rho\left[(1-\rho)^{-1}\sum_{s=1}^{\infty}(1-\rho)^s w_s + 2(1-\rho)^{-2}\sum_{s=2}^{\infty}(1-\rho)^s w_s + \ldots\right] =$$

$$\rho\sum_{s=1}^{\infty}(1-\rho)^s w_s\left\{(1-\rho)^{-1} + 2(1-\rho)^{-2}\left[1-\frac{(1-\rho)w_1}{\sum}\right] + 3(1-\rho)^{-3}\left[1-\frac{(1-\rho)w_1}{\sum}-\frac{(1-\rho)^2 w_2}{\sum}\right]+\ldots\right\} =$$

$$\rho\sum_{s=1}^{\infty}(1-\rho)^s w_s\left\{\left[(1-\rho)^{-1}+2(1-\rho)^{-2}+3(1-\rho)^{-3}+\ldots\right]-\frac{w_1}{\sum}\left[2(1-\rho)^{-1}+3(1-\rho)^{-2}+\ldots\right]-\frac{w_2}{\sum}\left[3(1-\rho)^{-1}+4(1-\rho)^{-2}+\ldots\right]-\ldots\right\} =$$

$$\rho\sum_{s=1}^{\infty}(1-\rho)^s w_s\left\{\sum_{k=1}^{\infty}(1-\rho)^{-k}k - \frac{w_1}{\sum}\sum_{k=1}^{\infty}(1-\rho)^{-k}(k+1) - \frac{w_2}{\sum}\sum_{k=1}^{\infty}(1-\rho)^{-k}(k+2)-\ldots\right\} =$$

$$\frac{1-\rho}{\rho}\sum_{s=1}^{\infty}(1-\rho)^s w_s - \rho\left\{w_1\left(\sum_{k=1}^{\infty}(1-\rho)^{-k}k+\sum_{k=1}^{\infty}(1-\rho)^{-k}\right)+w_2\left(\sum_{k=1}^{\infty}(1-\rho)^{-k}k+2\sum_{k=1}^{\infty}(1-\rho)^{-k}\right)+\ldots\right\} =$$

$$\frac{1-\rho}{\rho}\sum_{s=1}^{\infty}(1-\rho)^s w_s - \rho\left(\frac{1-\rho}{\rho^2}\sum_{k=1}^{\infty}w_k - \frac{1}{\rho}\sum_{k=1}^{\infty}kw_k\right) = \frac{1-\rho}{\rho}\sum_{s=1}^{\infty}(1-\rho)^s w_s - \frac{1-\rho}{\rho}\sum_{k=1}^{\infty}w_k + \sum_{k=1}^{\infty}kw_k.$$

Therefore:

$$a(\rho) \propto \sum_{k\geq 1}\left[EG(k)k - EL(k)k\right] = \frac{1-\rho}{\rho}\sum_{k=1}^{\infty}w_k - \sum_{k=1}^{\infty}kw_k = \frac{1-\rho}{\rho} - \sum_{k=1}^{\infty}kw_k.$$

In a similar way one can derive an adequate expression for *b(ρ)*:

$$b(\rho) \propto \frac{2-3\rho+\rho^2}{\rho^2} + \frac{2\rho-2}{\rho}\sum_{k\geq 1}kw_k + \sum_{k\geq 1}k^2 w_k.$$

# Figures

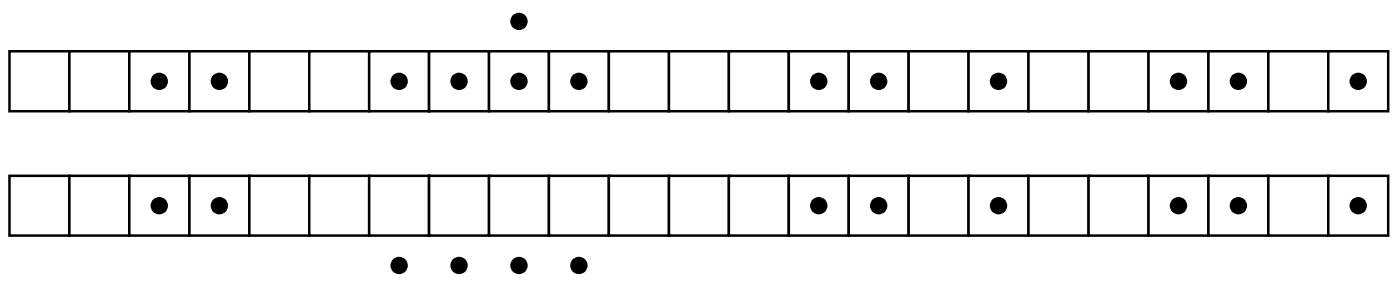

Fig. 1 Diagram presenting rules of the domino cellular automaton; an example of the avalanche of size four.

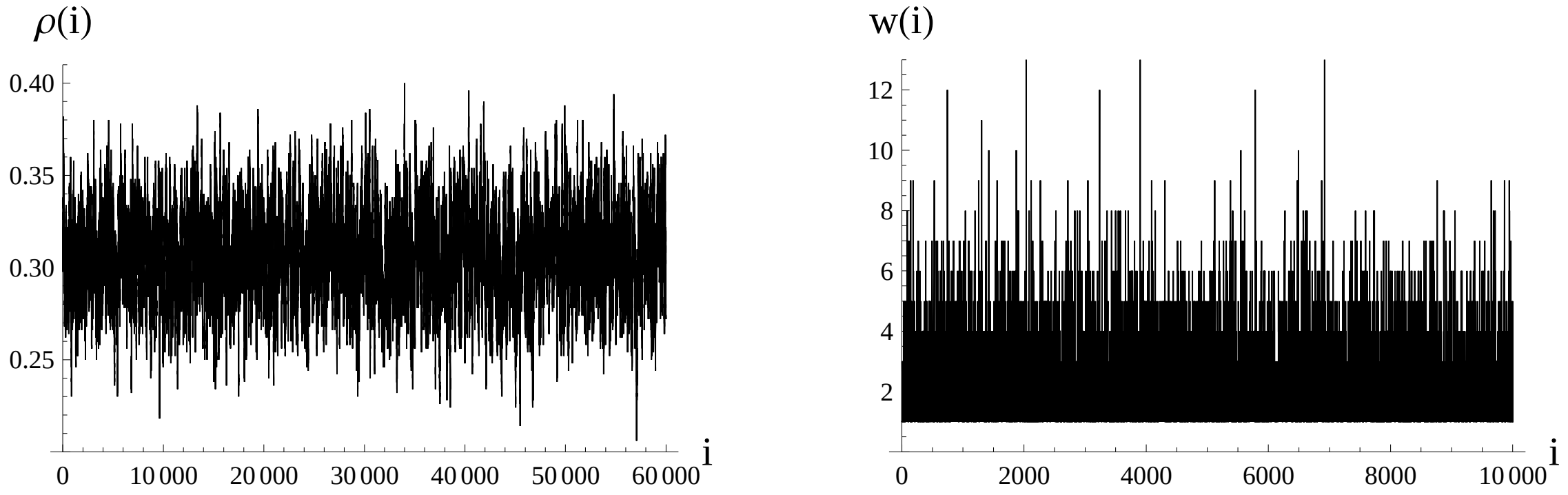

Fig. 2 The domino automaton at the quasi-equilibrium state: left – time series for density ρ*(i),* right – time series for avalanche sizes *w(i),*

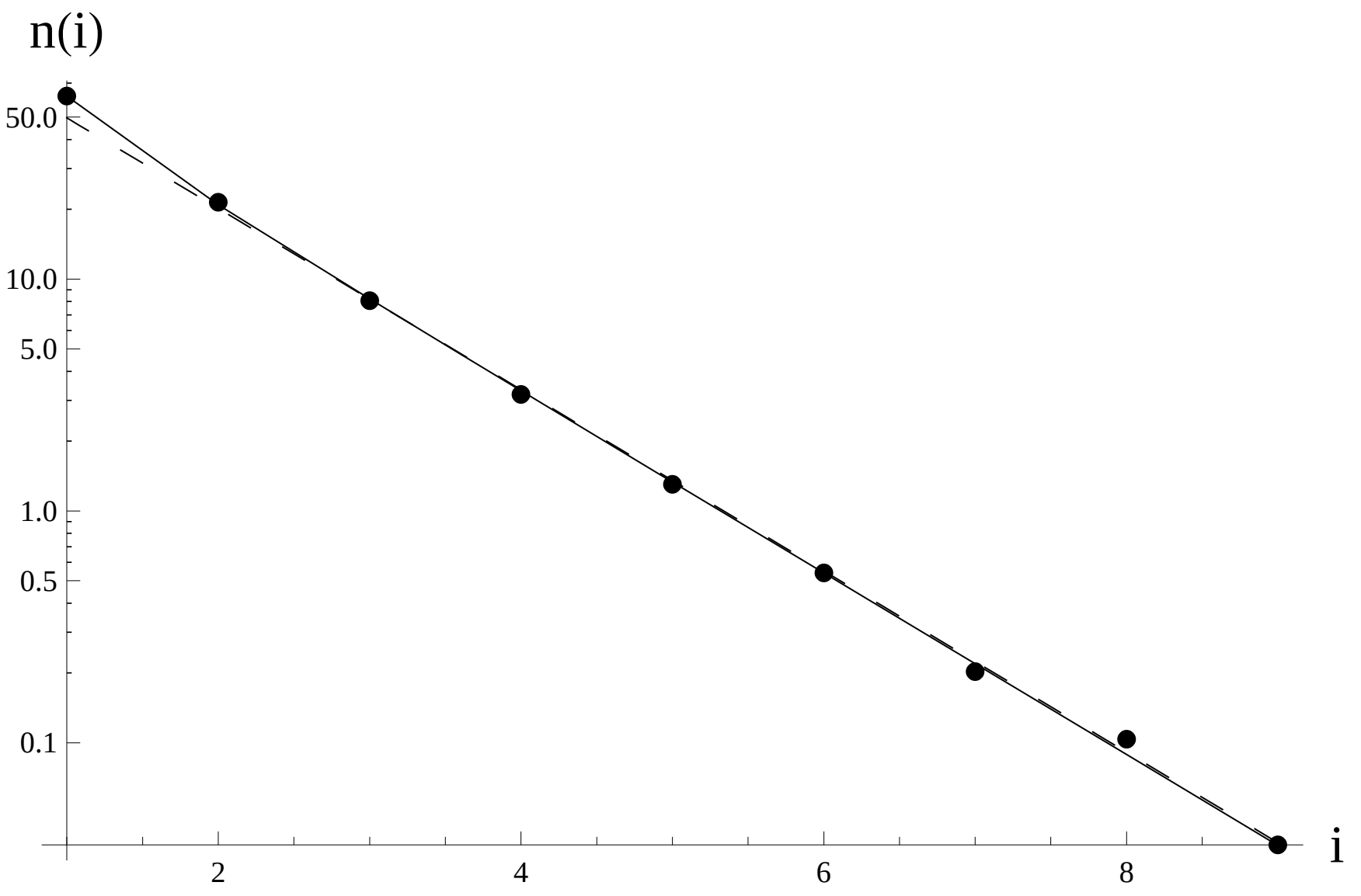

Fig. 3 Cluster size distribution function $n(i)$ at the equilibrium state; dots – simulation results, line – computed from equations (19)-(21), dashed line – exponential fit.

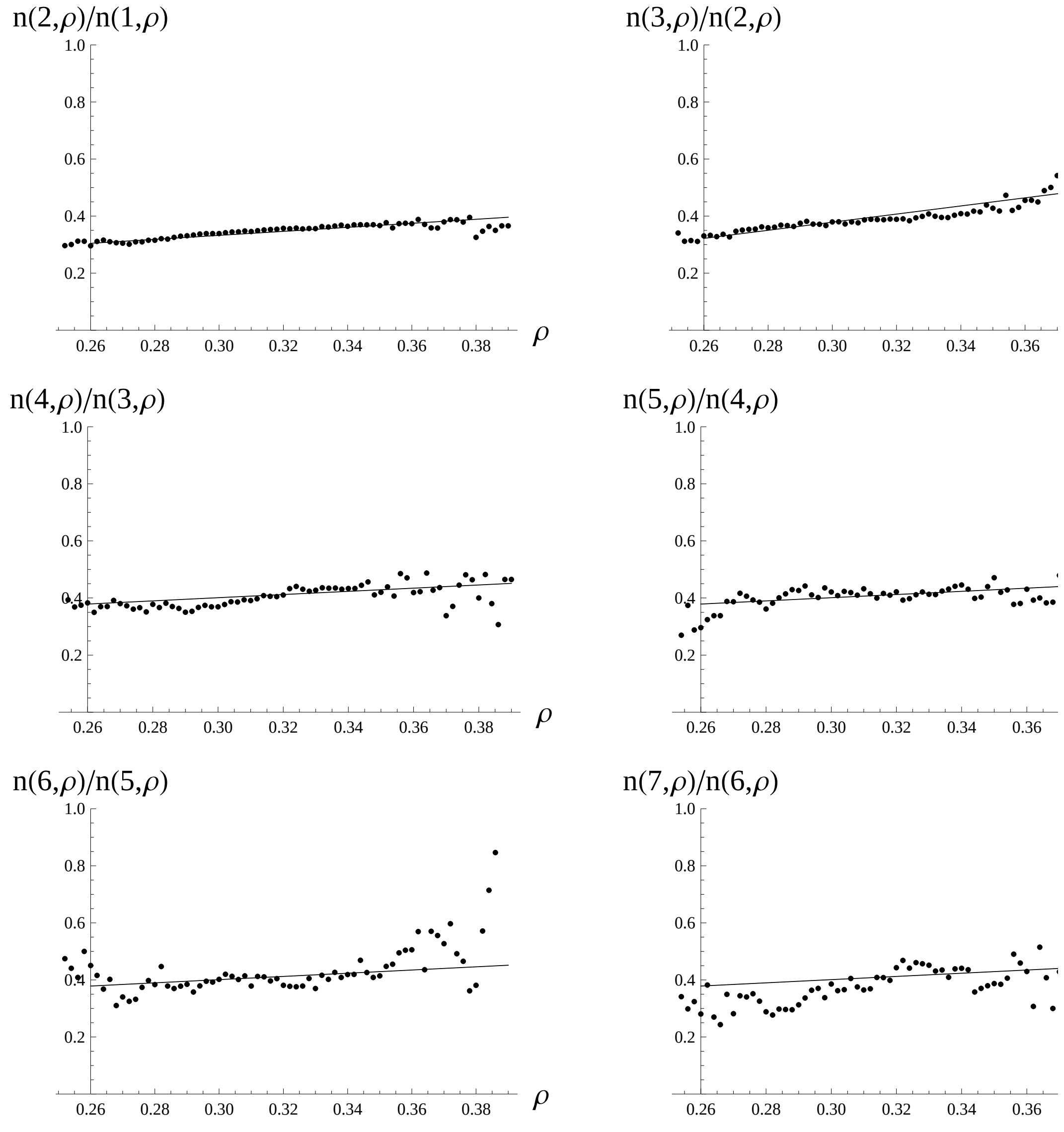

Fig. 4 Ratios $n_{k+1}/n_k$ , $k = 1, 2, \ldots, 6$, as functions of $\rho$; dots – simulation results, line – calculated from (19)-(21). They are nearly linear functions of $\rho$ in the range $0.23 < \rho < 0.4$.

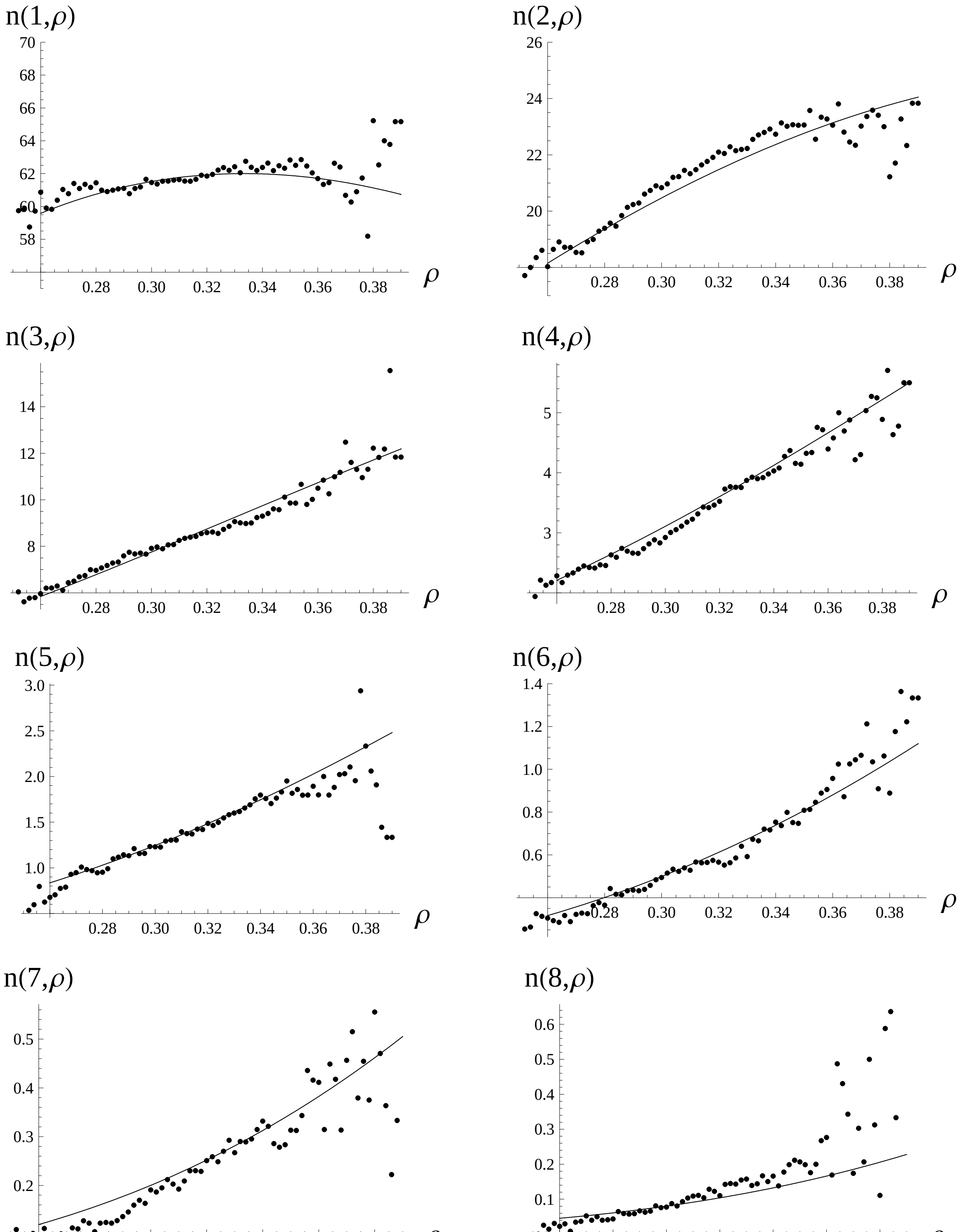

Fig. 5 Cluster size distribution $n_k(\rho)$, $k = 1, 2, \ldots, 8$, for deviation from the equilibrium state; dots – simulation results, line – calculated from (25) with (28), (29), (31), (33).

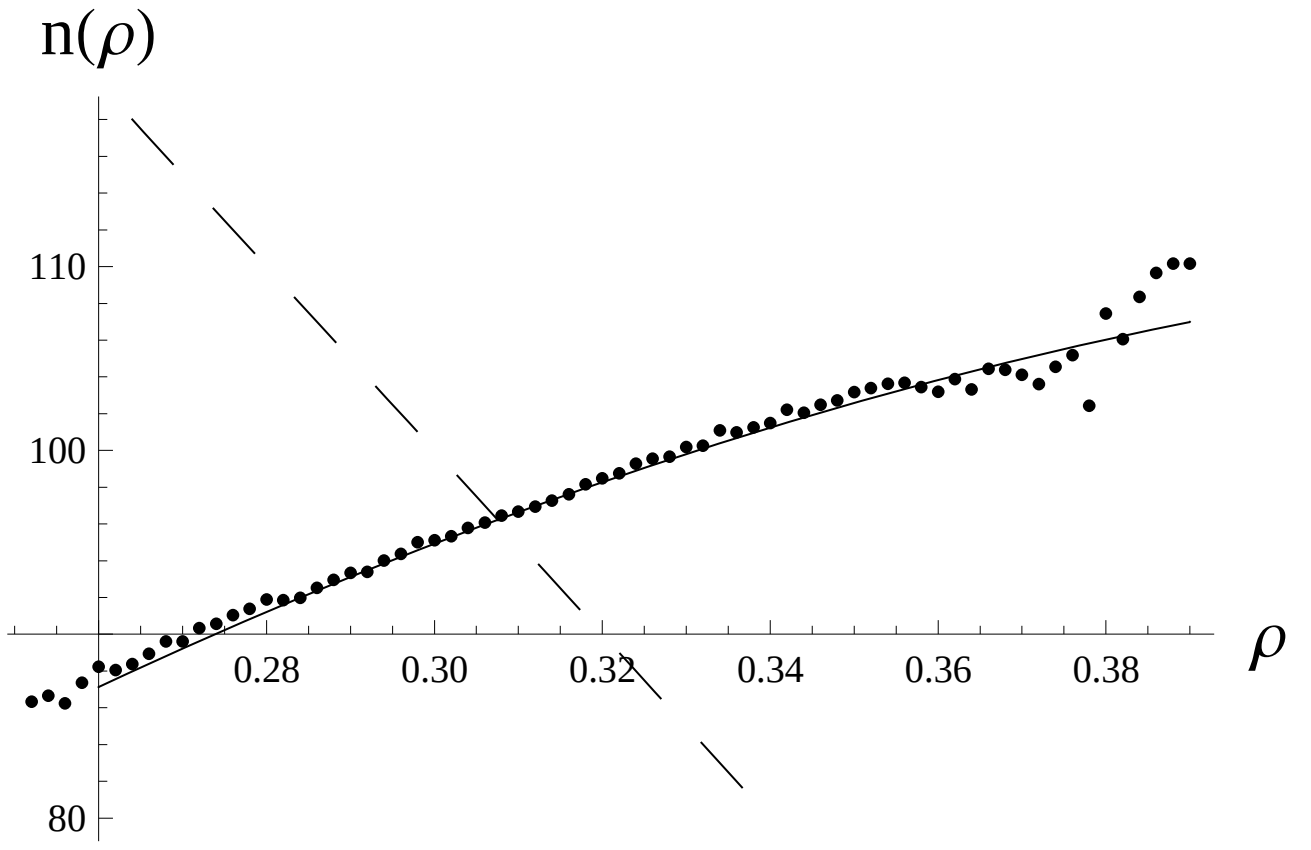

Fig. 6a The number of all clusters $n(\rho)$, dots – simulation results, continuous line – calculated from (34), dashed line – calculated from the equilibrium formula (22).

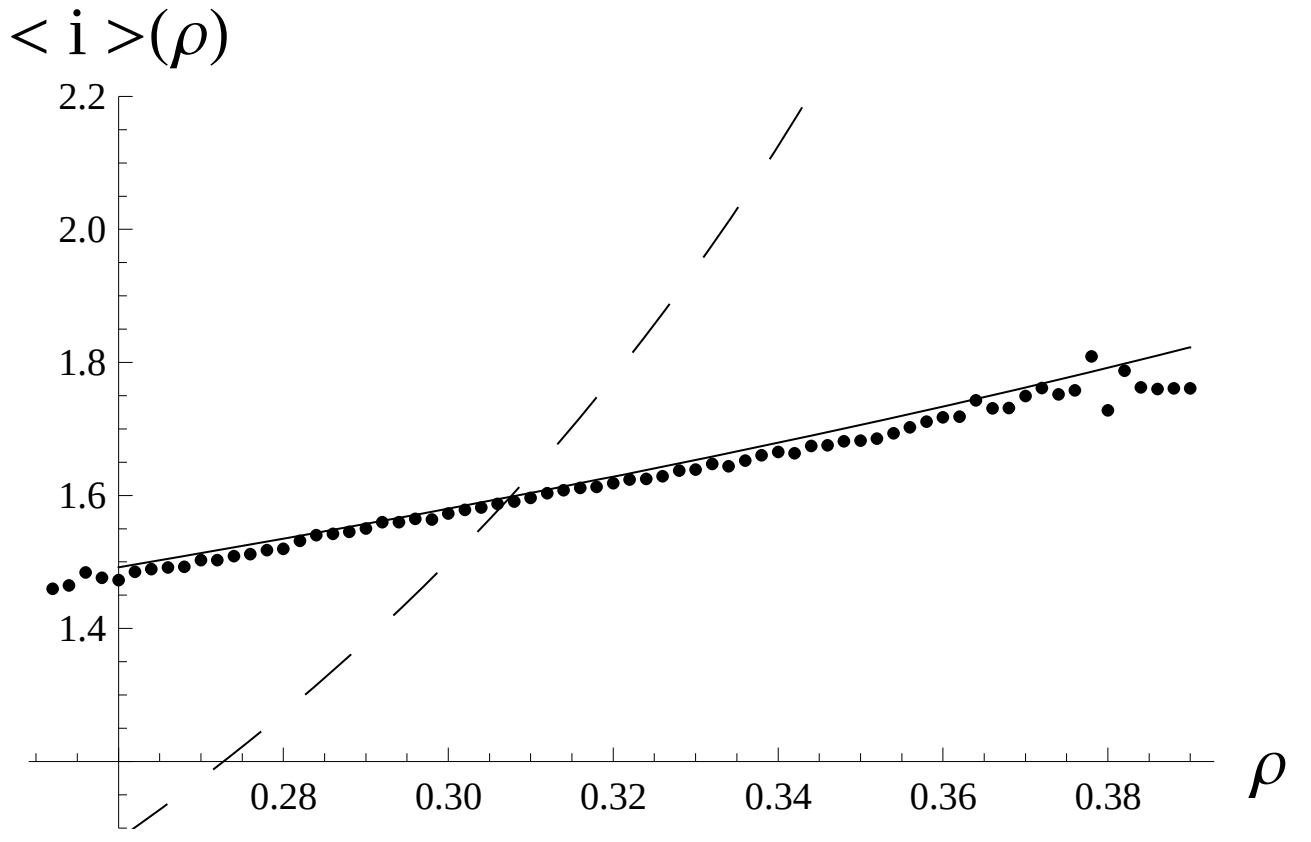

Fig. 6b The average clusters size $<i>(\rho)$, dots – simulation results, continuous line – calculated from (35), dashed line – calculated from the equilibrium formulas (7), (8).

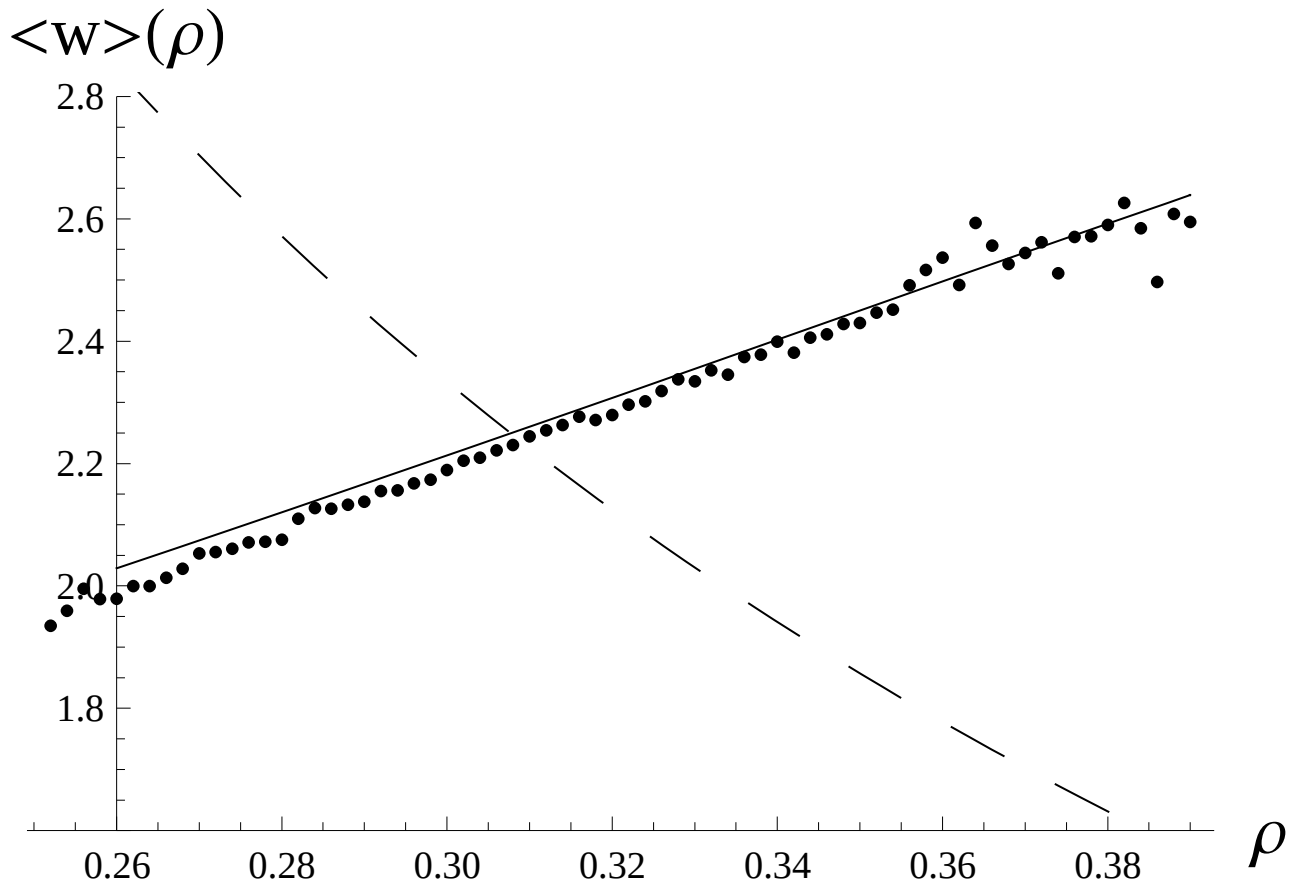

Fig. 6c The average size of avalanches $<w>(\rho)$, dots – simulation results, continuous line – calculated from (36), dashed line – calculated from the equilibrium formulas (7), (8).

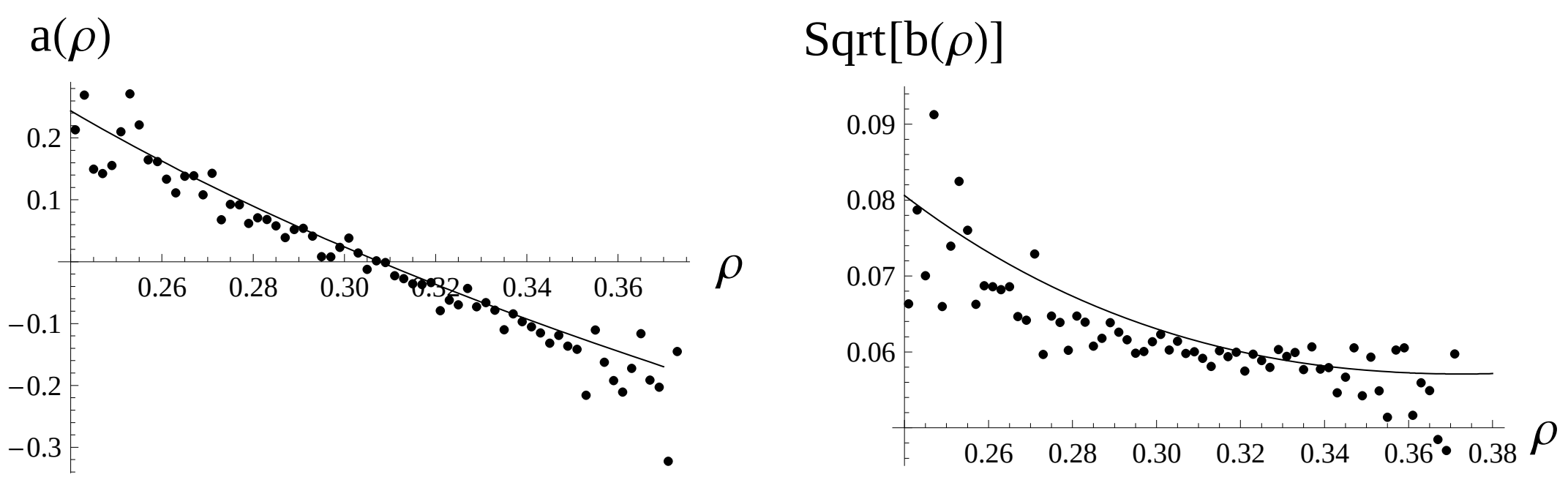

Fig. 7 Illustration of coefficients $a(\rho)$ and $\sqrt{b(\rho)}$ in the Ito equation; dots - reconstructed by the histogram method from the time series generated by the domino cellular automaton, lines – calculated from (45) - (48).

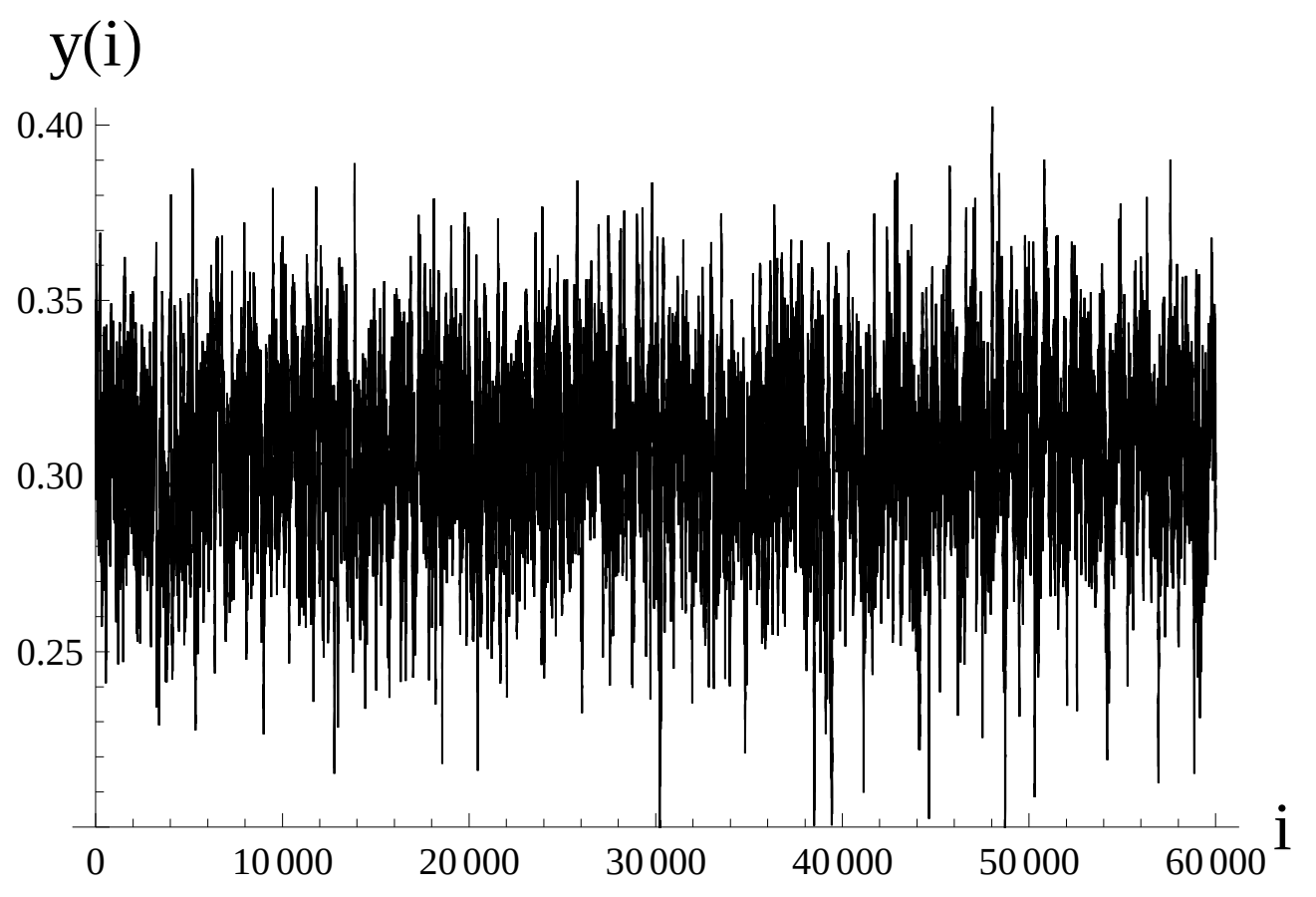

Fig. 8 Example of the time series generated by the reconstructed Ito equation (49) with (45)-(48).